\documentclass[preprint,superscriptaddress]{revtex4-1}
\usepackage{amsmath,amssymb}
\usepackage{graphicx}
\usepackage{epsfig}
\usepackage[sort&compress]{natbib}

\begin{document}

\title{Extended field-of-view in a lensless endoscope using an aperiodic multicore fiber}
\author{Siddharth Sivankutty}
\affiliation{Aix-Marseille Universit\'{e}, CNRS, Centrale Marseille, Institut Fresnel UMR 7249, 13013 Marseille, France}
\author{Viktor Tsvirkun}
\affiliation{Aix-Marseille Universit\'{e}, CNRS, Centrale Marseille, Institut Fresnel UMR 7249, 13013 Marseille, France}
\author{G\'{e}raud Bouwmans}
\affiliation{PhLAM CNRS, IRCICA, Universit\'{e} Lille 1, 59658 Villeneuve d'Ascq Cedex, France}
\author {Dani Kogan}
\affiliation{Department of Physics of Complex Systems, Weizmann Institute of Science, Rehovot, Israel}
\author {Dan Oron}
\affiliation{Department of Physics of Complex Systems, Weizmann Institute of Science, Rehovot, Israel}
\author{Esben Ravn Andresen}
\email{esben.andresen@ircica.univ-lille1.fr}
\affiliation{PhLAM CNRS, IRCICA, Universit\'{e} Lille 1, 59658 Villeneuve d'Ascq Cedex, France}
\author{Herv\'{e} Rigneault}
\email{herve.rigneault@fresnel.fr}
\affiliation{Aix-Marseille Universit\'{e}, CNRS, Centrale Marseille, Institut Fresnel UMR 7249, 13013 Marseille, France}

\begin{abstract}
We investigate lensless endoscopy using coherent beam combining and aperiodic multicore fibers (MCF). We show that diffracted orders, inherent to MCF with periodically arranged cores, dramatically reduce the field of view (FoV) and that randomness in MCF core positions can increase the FoV up to the diffraction limit set by a single fiber core, while maintaining MCF experimental feasibility. We demonstrate experimentally pixelation-free lensless endoscopy imaging over a 120~micron FoV with an aperiodic MCF designed with widely spaced cores. We show that this system is suitable to perform beam scanning imaging by simply applying a tilt to the proximal wavefront.
\end{abstract}
\date{\today}
\maketitle

\section{Introduction}
Imaging deep into biological tissue with light microscopy has been a holy grail for researchers over several decades~\cite{ntziachristos2010going}. Endoscopy overcomes a significant obstacle, light scattering, by delivering a probe deep into the region of interest. In order to be minimally invasive, miniaturization of the probe is an absolute must. A new class of imaging techniques, fiber based lensless endoscopes, achieve control of light at the distal end of the fiber with no additional opto-mechanical elements~\cite{CizmarNatComm2012,ChoiPRL2012}. There is a keen interest in these lensless endoscopes as they allow the miniaturization of the probe down to a few hundred microns, permitting minimally invasive imaging. In addition to conventional imaging, these lensless endoscopes have now been employed in nonlinear imaging such as two photon excited fluorescence~(TPEF) imaging~\cite{Andresen2013a,Sivankutty2016,Morales-Delgado:15}. This offers immense possibilities due to its intrinsic 3D resolved, label-free and molecular-specific imaging.

Multi-core fibers (MCF) are one of the leading contenders for the realization of lensless endoscopes. Traditionally, fiber bundles have been used in a configuration where each fiber core acts as an effective pixel sampling the object. This has two drawbacks: imaging is restricted to a single plane and pixelation artifacts arise due to the spacing between the fiber cores. With the emergence of wavefront shaping techniques, these issues have been overcome~\cite{Kim2014,Porat2016}. The fiber cores no longer directly sample the object but rather form effective 'pixels' of the composite wavefront emerging from the fiber. This wavefront can then be controlled from the proximal end of the fiber with spatial light modulators to generate and scan a focus in all three dimensions~\cite{Thompson2011,Andresen2013}. Image formation now is analogous to laser scanning microscopy and allows for pixelation-free and 3D resolved imaging at various distances from the fiber endface. However, there are consequences of the segmentation of the composite wavefront at the distal end of the fiber. This manifests as the appearance of multiple side lobes in addition to the generated desired focal spot. These side lobes limit the FoV attainable with MCFs to significantly small values~\cite{Andresen2013,Kim2015}. While the effect of the numerical aperture (NA) and the coupling between the cores on image formation has been extensively studied in~\cite{Chen2008, Stasio2015}, the geometrical arrangement of the fiber cores and its effect on the FoV has not been examined. Here we examine in detail the impact of these side~lobes in the context of lensless endoscopy, propose strategies to extend the FoV, numerically study the design parameters for MCFs and demonstrate experimental results with such a custom fabricated aperiodic MCF.

\section{Image formation in a MCF with uncoupled cores}
In this letter, we specifically focus on the case of a MCF with very weakly coupled cores and a raster scanning mechanism for image formation. This configuration has numerous advantages such as a tranmission matrix with no off-diagonal elements, increased robustness and a practically infinite memory effect~\cite{Andresen2013a}. Hence, imaging through a MCF is closer to conventional optical systems and we had earlier employed this advantage to demonstrate video-rate imaging at 11 fps~\cite{Andresen2013}. 

Generation of a focal spot in MCF based endoscopes can be understood as a coherent combination of multiple individual beamlets arising from the different cores of the MCF. When the phase differences between the individual beamlets are optimized for a particular point on the object plane, this manifests as a bright focal spot~\cite{Thompson2011,Andresen2013}. It is now clear that the electric field at an arbitrary axial plane from the exit of the fiber depends on the i) NA of the cores and ii) the arrangement of the cores on the fiber face.

Let us consider a MCF endface which consist of $N$ cores at locations $(x_{j},y_{j})$  and each single mode core gives an emitted field that can be well described by a gaussian envelope ~$\mathit{U}$.
%\begin{equation}
%  \mathit{U}(x+x_{j},y+y_{j}) = \mathit{U_0} \exp\Big( {-\frac{x^{2} + y^{2}}{2\sigma^{2}}}\Big)
%\end{equation}
The total field emanating from the MCF endface is
\begin{align}
  \mathrm{E(x,y,0)} = \mathrm{U(x,y)} \otimes \overbrace{\sum_{j=1}^{N} \delta (x-x_{j}) \delta (y-y_{j})}^\textbf{AF}
\label{eqn:AF}
\end{align}
We are particularly interested in the second term on the right-hand side of Eq.\ref{eqn:AF}, the array factor (\textit{AF}), which describes the geometric distribution of the fiber cores on the MCF. Even though the lensless endoscope operates typically in the Fresnel regime, let us initially consider the far-field of the MCF endface for a more intuitive case of the influence of the \textit{AF}. In the far-field, the composite electric field is proportional to the Fourier transform of the field at the fiber face 
\begin{align}
  \mathrm{E(k_{x},k_{y})} &= \mathcal{F}[ \mathrm{E(x,y)} ] \nonumber \\
    &= \mathcal{F}[  \mathrm{U}(x,y) \otimes  \sum_{j=1}^{N} \delta (x-x_{j}) \delta (y-y_{j}) ] \nonumber \\
%    &=  \mathcal{F}[\mathrm{U}(x,y)]\mathcal{FT}[\sum_{j=1}^{N} \delta (x-x_{j}) \delta (y-y_{j})] \nonumber \\
    &= \mathrm{U(k_x,k_y)} \sum_{j=1}^{N}\exp\Big(i({k_{x}x_{j} + k_{y}y_{j})}\Big)
\end{align}

And so, the intensity in the far-field is
\begin{align}
  \mathrm{I(k_{x},k_{y})} &= \mathrm{E} \cdot \mathrm{{ E^{\ast}}} \nonumber \\ 
&=  \mathrm{I_{\mathrm{1 core}}(k_{x},k_{y})} \mathrm{IF(k_{x},k_{y})}    
    \label{eqn:FF}
\end{align}

In real space, this is a convolution of $\mathit{I_\mathrm{1core}}$, the intensity pattern of a single core in the Fraunhofer regime and an interference term~\textit{IF}. When the cores are in phase, the far-field intensity pattern has bright regions where the \textit{IF} exhibits peaks and this is further weighted by the envelope of a single core, $\mathit{I_\mathrm{1core}}$. So far, we have not made any assumptions about any periodicity in the core positions $(x_j,y_j)$. In most cases during the fabrication of MCFs, the fiber preforms are periodically stacked. Considering the simplest 1D case with a pitch~$ \Lambda$, the \textit{AF} can be described as a \textit{comb} function and the \textit{IF} in the far-field can be written as 

\begin{align}
\mathrm{IF} &= \left|\mathcal{F}\Big[AF\Big]\right|^2 =\left|\mathcal{F}\Big[\mathrm{comb}\Big(\dfrac{x}{\Lambda}\Big)\Big]\right|^2  =\left| \dfrac{1}{\Lambda} \mathrm{\times comb}(\Lambda k_x)\right|^2  \
  \label{eqn:comb}
  \end{align}
The \textit{IF} manifests as a scaled version of \textit{AF}. Hence, for any periodic distribution of the cores, there will appear discrete side lobes at a spacing inversely proportional to the original pitch~$\Lambda$. Now beam scanning in the transverse plane is tantamount to shifting the peaks of the \textit{IF} in two dimensions by applying a tip/tilt to the composite wavefront. The FoV is thus given by the maximum radial displacement of this central peak before another side lobe becomes as intense as the original spot. It becomes clear that the entailed periodicity highly constrains the FoV. While a straightforward solution is to reduce the pitch $\Lambda$~\cite{Stasio2015}, this would directly result in a higher crosstalk between the cores, negating the key advantages of uncoupled cores listed earlier. Ideally, an \textit{IF} term that has only a single global peak is desirable and this corresponds to randomly arranging the cores such that there is no discernible pitch in the \textit{AF}. This can be understood by realizing that \textit{IF}, as defined in Eq.\ref{eqn:FF}, is equivalent to the autocorrelation of \textit{AF}; and the autocorrelation of a random function is a singly-peaked function. Thus, the intense side lobes in the case of a periodic \textit{AF} would now be replaced by a weak speckle background in the aperiodic MCF.
\begin{figure}[top]
  \centering
  \includegraphics[width=0.9\columnwidth]{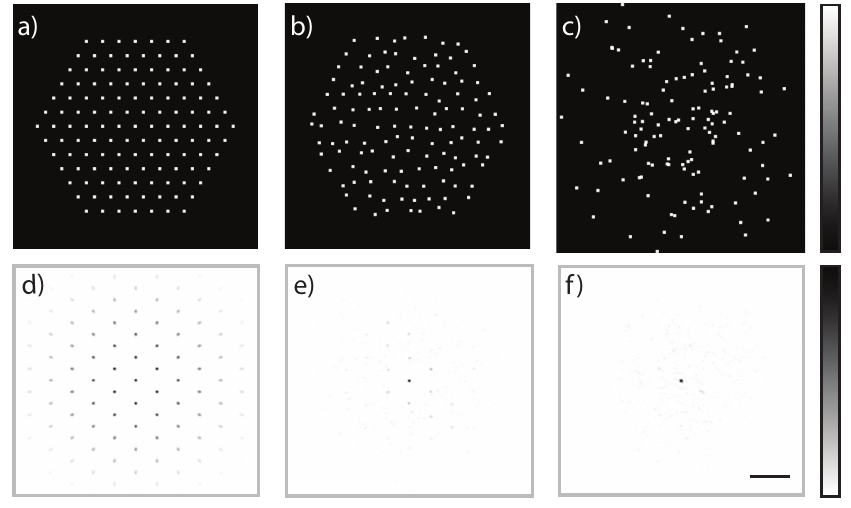}
  \caption{
Arrangement of the fiber cores in  a) a periodic lattice b)~a pseudo-random lattice with \textit{PR} = 0.22, c) a random arrangement and d)-f)  are their corresponding intensity distribution at a distance 500~$\mu$m from the fiber face. Scale bar - 50~$\mu$m. (See Visualization 1)
\label{fig:SatellitevsrPR}}
\end{figure}

\section{Design and fabrication of a novel aperiodic MCF}
While truly random arrangements are desirable, they are impractical as it would result in a very low packing fraction (cores per unit area) to avoid crosstalk and would not be realistically amenable to fabrication. We propose a novel way of introducing randomness in a MCF that is still practical to fabricate. A triangular lattice is chosen  as a starting point for its superior packing fraction. Each core of the MCF is offset radially from the master lattice by a fixed distance~\textbf{r}, and then rotated by a random angle about the original center. This generates a pseudo-random distribution of the cores over the MCF endface. This approach is beneficial on two accounts, i) such an arrangement is compatible with conventional fiber drawing techniques and, ii) a degree of control can still be exerted on the inter-core coupling by the design parameter $\dfrac{\mathbf{r}}{\Lambda}$. This ratio is referred to as the pseudo-random parameter~\textit{PR} in the rest of the manuscript.

For maximizing the FoV, we vary \textit{PR} and study the intensity distribution at a plane in the Fresnel regime using the angular spectrum method~\cite{SalehTeich}. We seek to minimize the side lobes of the system and this in turn allows raster scanning over a larger range. The parameters for the numerical study are chosen to be: fiber core size = 2~$\mu$m, $\Lambda$ = 20~$\mu$m and MCF diameter = 200~$\mu$m. In Fig.\ref{fig:SatellitevsrPR}, we depict core arrangments for 127 cores on (a) a periodic lattice, (b) a pseudo-random arrangement with \textit{PR}~=~0.22 and (c) a random arrangement constrained over a 300~$\mu$m region. Fig.\ref{fig:SatellitevsrPR}(d-f) represent their corresponding intensity patterns at a distance 500~$\mu$m away from the MCF endface. The periodic case  exhibits prominent side lobes ( Fig.\ref{fig:SatellitevsrPR}(d)), these are much less pronounced for the pseudo-random MCF (Fig.\ref{fig:SatellitevsrPR}(e))  and resemble a weak speckle background in the random case (Fig.\ref{fig:SatellitevsrPR}(f)).
 \begin{figure}[t]
  \centering
  \includegraphics[width=0.9\columnwidth]{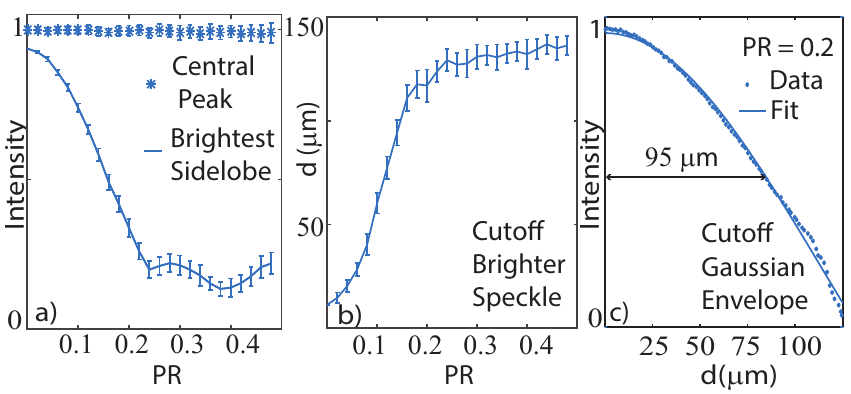}
  \caption{
Calculations - a) Relative brightness of the central peak (asterisks) and relative intensity of the brightest side lobe compared to the central spot (solid line) as a function of \textit{PR}. b)~Maximal radial displacement of the central spot before another speckle grain becomes equally bright with increasing \textit{PR} parameter and c) Relative intensity of the central lobe across the scan region.\label{fig:FoV_simulations}}
\end{figure}

Naturally, one wishes to obtain the maximum FoV and the smallest fraction of energy in the side lobes as possible. However, this needs to be balanced with a minimum core separation to ensure uncoupled cores. Hence, we perform numerical simulations to evaluate an optimal value for \textit{PR} as depicted in Fig.\ref{fig:FoV_simulations}. We first evaluate the effect of \textit{PR} on the intensity of the central lobe (asterisks) in Fig.\ref{fig:FoV_simulations}(a). It is straightforward to see that the intensity in the central lobe is not greatly perturbed (less than 2\%). This is in accordance with the simple far-field picture where the combining efficiency is dependent on the packing fraction, which remains the same in all the cases. However, when we evaluate the relative intensity of the brightest side lobe compared to the central lobe in Fig.\ref{fig:FoV_simulations}(a), the relevance of \textit{PR} becomes obvious. We see that even for small offsets, the intensity in the side lobes drop significantly and eventually plateaus for \textit{PR}~>~0.25. We note that the average intensity of the six brightest side lobes also follow a similar trend (data not shown). Values of \textit{PR} until 0.48 are investigated as for larger values, less control can be exerted over the separation of the cores during fabrication.

Furthermore, we examine the effective brightness of a spot across the FoV. A phase ramp is applied over the composite wavefront to scan the central lobe across the focal plane and we designate an empirical cut-off on the FoV when the desired central peak is no longer the brightest spot. 50 different realizations of core positions for each step of the \textit{PR} are generated and the resulting FoVs are examined in Fig.\ref{fig:FoV_simulations}(b). The maximum radial displacement plateaus at about 130~$\mu$m despite increasing the \textit{PR}~parameter. Here, we see that we are no longer limited by the periodicity of the cores but by the envelope term $\mathit{I_\mathrm{1core}}$, which is given by the divergence of a single core. We illustrate this graphically in Fig.\ref{fig:FoV_simulations}(c) in terms of the brightness of the central lobe across the FoV for \textit{PR}~=~0.2. Hence, we expect that any FoV limitation in our proposed pseudo-random arrangement is no longer from the memory effect or the periodicity of the fibers but from the diffraction envelope of a single core.
\begin{figure}[t]
  \centering
  \includegraphics[width=0.9\columnwidth]{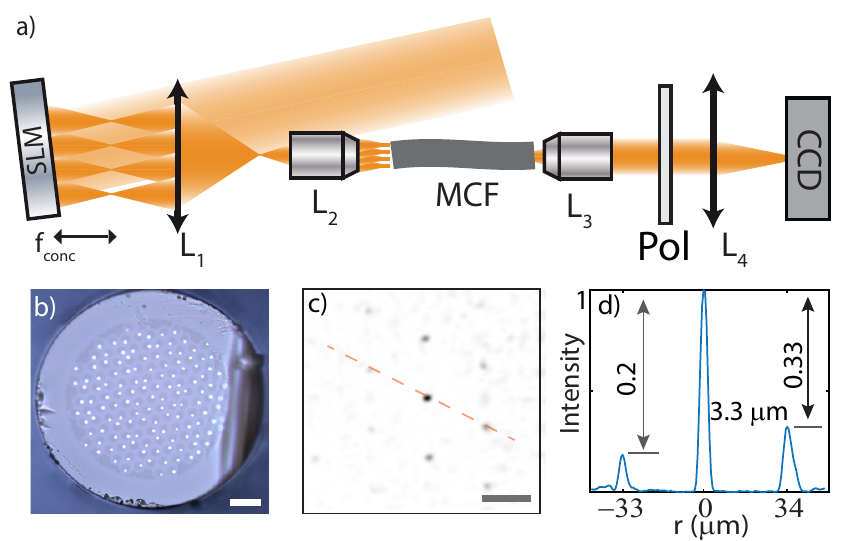}
  \caption{
a) Experimental setup. Annotations in text. b) An image of the fabricated MCF with a \textit{PR} of 0.22. Scale bar - 50~$\mu$m. c) An experimental measurement of the focal plane intensity at a plane 500~$\mu$m away from the distal end. Scale bar - 20~$\mu$m. d) Line profile (dashed line in (c)) the relative contrasts (0.2 and 0.33) and the FWHM of the central lobe (3.3~$\mu$m) are indicated. 
\label{fig:Expt}}
\end{figure}

In view of the calculations in Fig.\ref{fig:FoV_simulations} we choose PR $\approx$ 0.22 for the desired MCF. The diameter of an individual fiber~=~3.2~$\mu$m, NA~=~0.18 and a master pitch $\Lambda$~=~20~$\mu$m. This MCF has been made by the following original approach. A first stack of seven rods arranged in a triangular lattice was built into a silica tube. Six rods, including the central one, were in pure silica while the seventh (off-centered) was drawn from a  Ge-doped preform. This first assembly was then drawn into rods that were stacked onto a triangular lattice but with random orientations. This ensures that the distance between two neighboring Ge-doped cores takes random values between a minimum value (to ensure no significant coupling between cores) and a maximum value (to minimize the MCF diameter), these two values being related to the \textit{PR}. 

\section{Results}

A simplified view of the setup used to experimentally study the FoV is depicted in Fig.\ref{fig:Expt}(a). A liquid crystal SLM (X10468, Hamamatsu) is used to shape the wavefront of a femtosecond laser beam (1030 nm, 180 fs, t-pulse, Amplitude Syst\`emes) that enters the MCF. An array of focal spots is generated at a distance $\mathrm{f_{conc}}$ from the SLM and matches the relative core positions of the MCF. $L_1$ and $L_2$ represent a series of lenses that image this plane onto the proximal end of the fiber with a de-magnification of 55. Finally a telescope ($L_3, L_4$) images a plane 500~$\mu$m away from the distal end of the fiber onto a camera (CCD). A polarizer (Pol) placed ahead of the camera discards any effects of polarization scrambling in this non polarization maintaining MCF \cite{Sivankutty:16}. An inital calibration is performed and the intrinsic phase distortion introduced by the MCF is compensated. A more detailed description of this setup has previously been reported in~\cite{Andresen2013a}. Considering the use of femtosecond pulses, the MCF was 400 mm long and was held relatively straight to preclude any inter core group delay dispersion~\cite{Andresen2015a}. 

\begin{figure}[t]
  \centering
  \includegraphics[width=0.9\columnwidth]{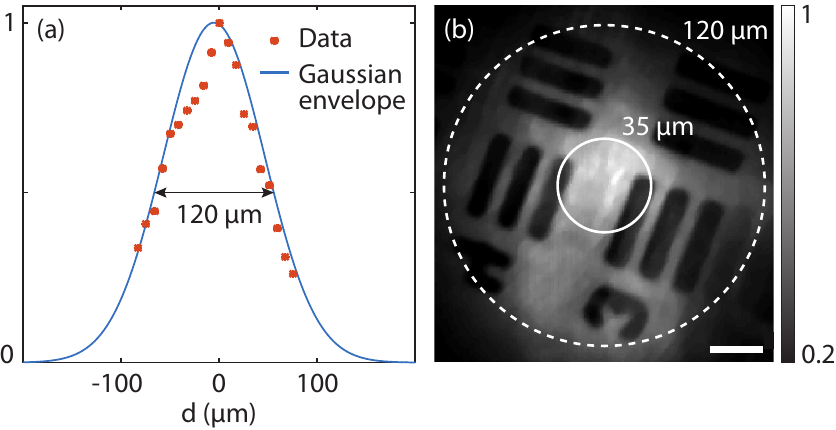}
  \caption{ a) Measured relative intensity of the central lobe across the FoV. b) Image of an USAF target (see text for details). The solid circle region (\O~= 35~$\mu$m) corresponds to the effective FoV in the case of a periodic arrangement and the dotted circle (\O~= 120~$\mu$m) for the aperiodic MCF. Scale bar - 20~$\mu$m.
\label{fig:Images}}
\end{figure}

Upon wavefront correction at the proximal end, we see the characteristic central lobe (FWHM = 3.3~$\mu$m) along with the six side lobes that are 35~$\mu$m apart as in Fig.\ref{fig:Expt}(c). The central spot, after the calibration, is about 97 times brighter than the average speckle background. In contrast with the periodic fiber, intensities of the side lobes are reduced at least by a factor 2.5, and a maximum reduction by a factor 8 is visible for the dimmest peak. Nevertheless, these side lobes are brighter than the average speckle background. Next, we apply a simple wavefront tilt at the proximal end, which translates to a shift of the central lobe at the distal end. This attests to the fact the cores have remained largely uncoupled. In Fig.\ref{fig:Images}(a), we evaluate the FoV by scanning the central lobe and as expected, the envelope factor limits the FoV to about 120~$\mu$m. This is in good agreement with the calculations presented in Fig.\ref{fig:FoV_simulations}, considering that there was less light injected into the outermost cores.

In the context of our earlier work~\cite{Andresen2013a}, the eventual endoscope is intended for nonlinear imaging like TPEF microscopy. In our present proof-of-principle fabrication, this aperiodic MCF does not have a high NA double clad, or a sufficient number of fibers to effectively collect the fluorescence signal. Hence, we perform an imaging experiment that experimentally simulates the TPEF contrast. We work in a transmission geometry with an USAF target, raster scan the focal spot and record the signal on a camera. In order to faithfully mimic a real TPEF experiment, we have to satisfy two criteria, i) the nonlinear dependence of the signal on the focal intensity. This is simulated by squaring the counts per pixel of the camera and ii) single pixel detection. This means no spatial information about the signal on the camera can be utilized for image reconstruction. This is achieved by summing up all the pixels. The effective two-photon signal on the $i^\mathrm{th}$ image pixel, $ I_{2p}(x_i,y_i)$, can be written as $ \sum I_n^2$ where $I_\mathrm{n}$ denotes the intensity on the $n^{th}$ pixel of the camera. Hence, any resulting image contrast has a direct quadratic relation to the focal intensity as in TPEF imaging. The resulting image (smoothed and background corrected for better visibility) is demonstrated in Fig.\ref{fig:Images}(b). The expected FoV of 120~$\mu$m (dotted circle) and the FoV that would result from a periodic arrangement,which is 35~$\mu$m (solid lines) are overlaid over the image for a visual comparison. While the absolute FoV might appear modest in comparison with conventional microscope objectives, one needs to consider the miniaturization of the probe (outer diameter~:~370~$\mu$m) which is an order of magnitude smaller than existing scanning endoscope systems. Taking this into account and considering the ratio of the FoV to the probe diameter, the aperiodic MCF compares favorably to most conventional microscopy systems. 

We have demonstrated for the first time pixelation free imaging through a sparse and widely spaced aperiodic MCF employing wavefront control. In addition to imaging, these concepts could also find applications in coherent beam combining using MCFs~\cite{Bourderionnet:11}. While the first results here have not employed epi-detection of the signal through the proximal end as in~\cite{Andresen2013a}, there is no significant difficulty in manufacturing such an aperiodic MCF with a high collection efficiency. Hence, we envision that these aperiodic MCFs would spur the development of miniaturized endoscopes and find wide applications in imaging systems.

\begin{acknowledgments}
The authors thank the following funding agencies:
Agence Nationale de la Recherche (ANR-11EQPX-0017,ANR-10-INSB-04-01,ANR 11-INSB-0006,ANR-14-CE17-0004-01), Aix-Marseille Universit\'e (ANR-11-IDEX-0001-02),Universit\'e Lille 1 (ANR-11-LABX-0007), Institut National de la Sant\'e et de la Recherche M\'edicale (PC201508),SATT Sud-Est (GDC Lensless endoscope),CNRS - Weizmann NaBi International Laboratory.
We thank R\'emi Habert and Karen Delplace for technical assistance.
\end{acknowledgments}

%}

\end{document}